# Understanding and Modeling AI-Intensive System Development

Luigi Lavazza

Dipartimento di Scienze Teoriche e Applicate
Università degli Studi dell'Insubria
Varese, Italia
luigi.lavazza@uninsubria.it
ORCID 0000-0002-5226-4337

Sandro Morasca

Dipartimento di Scienze Teoriche e Applicate
Università degli Studi dell'Insubria
Varese, Italy
sandro.morasca@uninsubria.it
ORCID 0000-0003-4598-7024

*Abstract*—Developers of AI-Intensive Systems—i.e., systems that involve both "traditional" software and Artificial Intelligence—are recognizing the need to organize development systematically and use engineered methods and tools. Since an AI-Intensive System (AIIS) relies heavily on software, it is expected that Software Engineering (SE) methods and tools can help. However, AIIS development differs from the development of "traditional" software systems in a few substantial aspects. Hence, traditional SE methods and tools are not suitable or sufficient by themselves and need to be adapted and extended.

A quest for "SE for AI" methods and tools has started. We believe that, in this effort, we should learn from experience and avoid repeating some of the mistakes made in the quest for SE in past years. To this end, a fundamental instrument is a set of concepts and a notation to deal with AIIS and the problems that characterize their development processes.

In this paper, we propose to describe AIIS via a notation that was proposed for SE and embeds a set of concepts that are suitable to represent AIIS as well. We demonstrate the usage of the notation by modeling some characteristics that are particularly relevant for AIIS.

*Keywords*—Artificial Intelligence (AI), Software Engineering (SE), development process, process models, modeling notation

## I. INTRODUCTION

The use of Artificial Intelligence (AI) techniques is increasingly spreading in many application fields in a pervasive fashion. AI techniques are usually used to derive models for classification, prediction, optimization, etc. These models are very often used as components for building "AI-Intensive" systems (AIIS[1]). These systems also include a large "traditional" software part that manages I/O, interface with users, possibly connects with other systems, performs administration tasks, etc.

A large number of methods, techniques and tools have been specifically conceived and implemented to support traditional software development. This large corpus is the basis of the Software Engineering (SE) discipline. However, AIIS have specific characteristics that make them quite different from more conventional software systems under several respects. For instance, a development focus of AIIS is on managing data (such as how to store big and evolving data, how to check the quality of data, how to create knowledge from data, etc.). Also, the specifications of an AIIS application may not be fully known before one delves into the data and investigates what can be obtained out of them.

Methods and tools are still lacking to address the development of AIIS, i.e., the integrated development of AI models and the software applications that use them. The proposal of these methods and tools can benefit from the amount of previous experience on the quest for effective SE methods and tools, to learn from previous mistakes and avoid making them again.

In this paper, we propose to describe AIIS via a notation that was proposed by Jackson for traditional software systems [1]. This notation helps reason about the properties of AIIS and their similarities and differences with respect to conventional software systems. For instance, it is acceptable that an AIIS does not always return a correct result, while this would not be acceptable, at least in principle, in conventional software systems. Thus, even basic qualities like correctness need to be rethought, modeled, adapted, and measured in possibly different ways than in traditional SE.

---

[1] We use the "AIIS" acronym for both the singular and the plural forms of "AI-Intensive System."

The remainder of the paper is organized as follows. Section II describes the notation and how it can be used for traditional software development. Section III shows how the notation can be used when it comes to representing AIIS. We investigate how qualities of interest in AIIS can be defined in a possibly quantitative fashion in Section IV. Relevant related work is concisely summarized in Section V and conclusions and an outline for future work are presented in Section VI.

## II. BACKGROUND

### A. Basic Concepts and Notation

The proposed notation is based on the following formula

$$E, S \vdash R \qquad (1)$$

which was introduced by Jackson [1]. In Formula (1), *R* indicates the stakeholder's requirements, i.e., "conditions over the phenomena of the environment that we wish to make true by installing the machine" while *E* expresses "conditions over the phenomena of the environment that we know to be true irrespective of the properties and behavior of the machine" [1]. *S* is the specification of the machine, expressed in terms of the phenomena that are shared by the environment and the machine. In practice, *S* is specified in terms of I/O elements: in Jackson's terminology, phenomena controlled by the environment and visible to the machine—i.e., inputs— and phenomena controlled by the machine and visible to the environment—i.e., outputs.

So, Formula (1) is an entailment that reads: if the environment in which the machine is located behaves as specified in *E* and the specifications *S* are satisfied by the machine and the environment, then requirements *R* are satisfied. The logical entailment $A \vdash B$ states that from assuming A we can prove B; hence, entailment is often called provability.

The level of formality of Formula (1) can vary. In fact, it depends on the formality of the descriptions *E*, *S* and *R*. It is possible to provide informal, text-based descriptions, as well as completely formal specifications, depending on the intended use of these descriptions. Of course, if *E*, *R* and *S* are described formally, it is possible to prove whether the truth of *R* descends from the truth of *E* and *S*, while informal descriptions allow only for argumentations on whether the truth of *R* descends from the truth of *E* and *S*.

We use the concepts and notations originally conceived by Jackson, which were later described as a reference model for software requirements by Guenther et al. [2].

### B. Modeling Traditional Software Development

Software development is needed when

$$E \nvdash R \qquad (2)$$

since no software would be required if $E \vdash R$, i.e., if requirement *R* were already "spontaneously" satisfied in the environment.

When building traditional software, we rely on the description *E* of the environment. However complex the environment, we assume that the knowledge of the environment expressed via *E* is sufficient to describe the portion of the environment's behavior of interest to the stakeholders. *E* is obtained by observing the actual behavior of the environment, or by considering cause-effect laws that others have described, based on their own observations.

In general, knowing *E* and *R*, it is possible to conceive a machine that

1) Works as specified by *S*;
2) When introduced in the environment, causes *R* to be satisfied, i.e., $E, S \vdash R$.

Thus, one needs to build a "machine" that behaves as specified by *S*. This "machine" is in general obtained by taking a ready-to-use "platform" *M* and configuring it via some program *P* built specifically to satisfy *S*.

$$E, P, M \vdash R \tag{3}$$

Usually, the platform includes hardware, an operating system, some middleware and software utilities, to minimize the effort required to build program *P*.

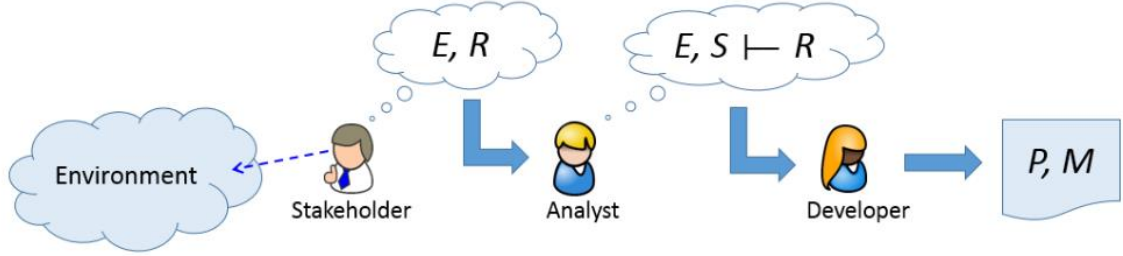

Fig. 1. A schematic view of traditional software development.

Figure 1 schematizes traditional software development. The delivered *P*, along with the selected platform *M*, is supposed to successfully implement *S*, i.e.,

$$P, M \vdash S \tag{4}$$

However, such entailment is hardly ever subject to formal proof. Developers typically use various techniques, such as Verification & Validation techniques like software testing and inspections, to provide evidence on the truthfulness of Formula (3). For the sake of precision, proving (4) is the objective of verification, while proving (3) is the objective of validation.

In general, several definitions of *S* satisfy Formula (1) and, similarly, several pairs ⟨*P, M*⟩ satisfy *S*. The choice of *S*, *M*, and *P* is usually mainly guided by economic considerations, i.e., *S* should be implemented in reasonably short time and with a sustainable effort.

## III. DESCRIBING AIIS

AIIS, with special reference to systems that include a component obtained via Machine Learning (ML) or similar techniques, can be described as the composition of

- the platform *M*;
- a model *Mdl* that supplies artificial intelligence to the system;
- the "traditional" software *P*, in which *Mdl* is embedded.

Therefore, we expect that an AIIS satisfies users' needs if

$$E, M, Mdl, P \vdash R \tag{5}$$

Now, let us consider how *Mdl* is obtained. In general, *Mdl* is built by analyzing data that are produced by the environment. For instance, a model that detects and classifies tumors can be obtained by applying data analysis techniques to tomography scan images. Similarly, a chess playing model can be obtained by analyzing a set of chess game recordings.

The need for *Mdl*, and AI in general, comes from the inability to describe environment *E* satisfactorily. AI models would not be needed if we were able to completely understand the relationships between the graphical characteristics of tomography scans and the characteristics of tumors, or the best move in any given chessboard situation.

There is an important difference between traditional systems and AIIS. In traditional systems, *E* describes "all we need to know" about the considered portion of the environment. It is in general possible to describe *E* completely and precisely. Of course, mistakes are always possible: for instance, the famous Ariane 5 disaster was due to an inadequate description of the range of possible horizontal speeds of the rocket. Nonetheless, a good domain expert should always be able to provide *E* that correctly describes the environment, in principle.

Based on *E* and *R*, it is possible to write *S*, which contains all the information needed to select *M* and build *P*.

This is not the case with AIIS, which are built under the following circumstances.

- We have an incomplete understanding of the relationships that link the phenomena in the environment with the requirements. Specifically, we have a limited understanding of the cause-effect relationships in the environment: because of this limited comprehension, we are generally not able to devise actions that modify the environment's behavior so that requirements are always certainly satisfied. For instance, given a chessboard configuration, we are not able to say if there is a move that may bring us to win the game regardless of what the opponent may do.
- Ideal requirements are known as usual. For instance, we would like a diagnostic system that always correctly classifies the nature of a tumor, based on the available tomography scan images; similarly, we would like a chess playing system that always wins against any opponent (including other AIIS).
- We are aware that if we express the aforementioned ideal requirements, it will be practically impossible to achieve them. In practice, we would probably be happy with a diagnostic system that performs as well as a good physician, and with a chess playing system that is able to win most times with the best players.
- We do not know—except approximately, maybe— the phenomena that the environment may generate. Usually, we are able to describe very general situations (e.g., the set of all legal chessboard configurations), while it is much more difficult to describe the subset of situations that are likely to occur (e.g., chessboard configurations that can be reached during real games).
- We have data (possibly in very large quantity), which record the observed behavior of the environment. As usual, data represent a sort of extensive knowledge that we can exploit. We name these data $D_E$. Accordingly, we assume that *E* describes—among other things—the ability of the environment to produce $D_E$.

We can represent the role of AIIS via Formula (5), which shows what we would like to achieve, ideally. To understand and model the construction of AIIS, let us observe that the environment we have modeled is able to produce a large quantity of data (each instance representing a specific behavior), and we usually have observed only a relatively small (though possibly large, in absolute terms) subset of behavior instances. For example, we have a large collection of images produced by tomography devices. Similarly, we have thousands of chess game recordings. However, the collections of scans or game records we own are very small subsets of the set of all possible images or games. We can represent this situation as follows:

$$E \vdash D_{tot} \supset D_E \qquad (6)$$

where $D_E$ is the set of data we actually own that represent the environment, while $D_{tot}$ is the entire set of data the environment could produce.

A *Mdl* model is built based on the available data, i.e.,

$$D_E, M, P, Mdl, \vdash S_D \qquad (7)$$

Recall that specifications S (or SD in (7)) represent the relationships between the part of the environment that is visible to the system and the part of the system that is visible to the environment. In general, these phenomena are termed "inputs" and "outputs," respectively. Therefore, Formula (7) says that, when it receives an input belonging to DE, our system will yield the output as specified in SD.

*Mdl* is built based only on $D_E$, while *E* can produce also different data, as shown in Formula (6). Thus, we have in general that

$$E, M, P, Mdl \not\vdash R \qquad (8)$$

$$E, S_D \not\vdash R \qquad (9)$$

In fact, it is possible that *E* yields some input data belonging to the difference $D - D_E$ that are not properly handled by *Mdl*.

This problem is mitigated by the fact that *Mdl* is built in a way to achieve a behavior that is acceptable also in many cases that are not represented in $D_E$. In fact, the main goal of AI is to *learn* from the available data, so that the resulting AIIS can behave acceptably also when dealing with unprecedented situations. Our diagnostic system should be able to correctly classify a large number of tomography scan images that do not belong to $D_E$. Similarly, we expect that a chess playing AIIS is able to respond effectively to a move that did not appear in $D_E$. This means that there exists D' such that

$$|D'| \gg |D_E| \tag{10}$$

$$D' \cap D_E \neq \emptyset \tag{11}$$

$$D', S_D \vdash R \tag{12}$$

In other words, there is a set of data (corresponding to manifestations of the environment) D' that includes a (large) part of $D_E$, is much larger than $D_E$, and satisfies requirements R.

However, we know that AIIS will not be able to *always* satisfy R. For instance, we do not expect that a diagnostic AIIS *always* provides a correct response. It should be noted that it would be very easy to build a system that makes $D_E, S_D \vdash R$ true—but then fails often when the input data do not belong to $D_E$. Such AIIS would be hardly useful: what we need is a system that is able to behave correctly when dealing with D' described above.

Therefore, an AIIS in general successfully satisfies R for a—hopefully large—subset $D_{ES}$ of $D_E$, but fails for some other—hopefully very small—subset $D_{EF}$. This situation can be described as follows:

$$D_E = D_{ES} \cup D_{EF} \wedge D_{ES} \cap D_{EF} \neq \emptyset \tag{13}$$

$$D_{ES}, S_D \vdash R \tag{14}$$

$$D_{EF}, S_D \nvdash R \tag{15}$$

In general, $|D_{EF}| \ll |D_{ES}|$. Ideally, $|D_{EF}| = 0$, but this hardly ever happens.

It should be noted that the situation described by Formulae (13), (14), and (15) holds for traditional systems as well, especially if you substitute *S* with the implementation. In practice, there are always some input data that cause real systems to fail. However, in traditional development, this is just a consequence of errors in the specification or implementation of a software system: in principle, we know what should be done, but we fail to do it. Instead, in AIIS, the fact that $D_{EN}$ is not empty is an intrinsic, unavoidable characteristic. Especially in the case of ML and Deep Learning AIIS, we build systems that are expected to provide correct answers in most cases, but not necessarily all.

Different techniques can be applied to derive a *Mdl* that addresses the problem summarized by Formula (7). Therefore, we may achieve different $S_D$, which split $D_E$ into different pairs of subsets $D_{ES}$ and $D_{EF}$. Here, we have a quite important difference with respect to traditional software development. In fact, we can build different systems that implement *S* (as in Formula (4)), hence satisfy *R* (as in Formula (1)), but the choice of *M* and *P* is usually driven by considerations concerning the available technology, the time and effort needed to build *P*, etc. Instead, when dealing with AIIS, the choice of *Mdl* in Formula (7) is usually driven by the desire to make $D_{EF}$ as small as possible.

## IV. REQUIREMENTS AND SOFTWARE QUALITIES

Requirements have traditionally been divided into functional and non-functional ones in SE [3]. Non functional requirements are typically characterized as software qualities. It is inherent to the diversified nature of software application domains that different applications have different emphasis on different qualities. For instance, efficiency is typically a much more important quality in embedded software applications than usability, while the converse may be true for personal productivity applications.

AI and ML can be applied to a large number of vastly diverse application domains, each one characterized by specific qualities of interest. However, there are a few common qualities that are important in AI- and ML based applications, regardless of the specific domain. As often happens, these qualities may be antithetical, i.e., optimizing one may lead to suboptimal levels in others.

We need a comprehensive view on these qualities, i.e., we need quality models for AIIS. These quality models should be as "quantitative" as possible, more so than existing quality models, e.g., SQUARE 25000 [4], for two reasons.

1) Quantitative quality models make it possible to have quantitative requirements, at least to the extent possible, so their satisfaction can be truly checked. Requirements like, for instance, "the application must be reliable" and "the application needs to be highly usable" clearly ring hollow, just like requirements such as "the new version of the system needs to be more reliable" and "more usable" than the previous version. All of these requirements are untestable, because no precise definition is provided of what "reliable" and "usable" mean and how they are quantified. Lacking those precise definitions, the satisfaction of those requirements can be checked only on very subjective terms, which is scientifically unsatisfactory and inconvenient from a practical point of view.

2) AI and ML are based on largely quantitative techniques, rooted in Statistics and other mathematical concepts. The quality of AI and ML algorithms is evaluated in very quantitative terms, via several different performance[2] metrics. Thus, it would be quite strange to have non-quantitative requirements for AIIS, even though the SE world has so far largely got away with non-quantitative requirements.

The topic needs to be addressed in much greater detail and a broad consensus needs to be ultimately built around the important qualities for AIIS. Here, we discuss a few qualities and issues that can be interesting for AIIS.

The discussion of AIIS qualities that follows is expected to help in better understanding and organizing AIIS testing activities. Being able to define better the properties that are the object of testing seems to be increasingly needed, as AIIS testing (and ML testing, specifically) are gaining interest in the scientific community [5].

*A. Functional Correctness*

Functional correctness should go beyond what is commonly taken into account as correctness. For instance, the idea that AIIS is correct because it implements correctly some ML algorithm may be unsatisfactory in practical applications. What matters is that the performance of the AIIS is above some specified level for a set of selected performance metrics of interest. As a parallel to medical tests, when it comes to building a testing device for some pathogen, what is usually considered important is that the device has Sensibility and Sensitivity above some specified thresholds. The method used for the detection of the pathogen is clearly fundamental, but its correct implementation by means of the device is not a sufficient condition for the usefulness of the device— and not even a necessary one, in principle. All that truly matters is that the levels of Sensibility and Sensitivity are above some specified thresholds.

Likewise, in AIIS, one could say that it is not so important anymore whether the software correctly implements a given algorithm, but, rather, whether the results are practically acceptable. As a totally extreme example, an AIIS may not correctly implement a given algorithm (or, as a less extreme example, it may set the hyperparameters at different values than the ones that were actually intended), but it may nonetheless provide better performance results than those that would be obtained via a functionally correct implementation. This shows the importance of having quantitative, testable requirements, instead of generic requirements, such as "precision should be satisfactory."

Taking into account performance metrics introduces a few additional challenges to the usual way of checking correctness.

It seems sensible that the AIIS must satisfy performance requirements at least on the training set. However, the performance of the AIIS is then checked via test sets and it would be unrealistic to expect that the performance requirements must be satisfied on every possible test set. Thus, constraints should be set on the performance of the AIIS in terms of, for instance, average performance and standard deviation, or in terms of the probability of satisfying the performance requirements.

---

[2] Here the term "performance" is used, as usual in the ML domain, to indicate the accuracy of classifications, predictions, etc. yielded by AI software. Instead, in the SE domain "performance" is usually related to efficiency in resource consumption.

After all, software is hardly ever released with perfect certitude that it is correct. In many cases, it is released even though a set of "known bugs" has not been resolved.

The AIIS is then put to operational use and its performance needs to be constantly monitored. There need to be rules in place to "maintain" the AIIS whenever it no longer appears to satisfy the performance requirements. For instance, this is what happens during quality control in manufacturing. A series of artifacts is examined, one after the other. The system is considered as functioning correctly until the evidence to the contrary increases past a specified level. This may be due to several reasons, including a concept shift. AIIS will learn from its mistakes, i.e., by retraining the ML part based on the newly acquired data, but that may not be enough.

*Specifying Functional Correctness*

Formulae (13), (14), and (15) provide the basis to specify functional correctness of AIIS quantitatively. Among the many performance metrics that have been proposed [6], "accuracy" (*acc*) can be easily computed using $D_{ES}$ and $D_{EF}$

$$acc = \frac{|D_{ES}|}{|D_{ES}|+|D_{EF}|} = \frac{|D_{ES}|}{|D_E|} \tag{16}$$

We may state that *Mdl* has been correctly implemented when, applied to $D_E$, it yields *acc* > *t*, where *t* is the minimum acceptable performance.

Here we use *acc* for convenience: any other performance metrics (e.g., precision, recall, F1, AUC, etc.) can be used to evaluate functional correctness. The choice of performance metric depends on the goals of the AIIS stakeholders.

The threshold value for a selected performance metric depends on the stakeholders' goals too and, in come cases, can be set based on the observation (and measurement) of the environment. For instance, we may compute the diagnostic accuracy of a set of doctors and look for a *Mdl* whose *acc*, as defined in (16), is not worse than the average accuracy of doctors.

Assessing the functional correctness of *Mdl*'s implementation is a first, necessary step. However, we are likely very interested in the correctness of *Mdl in use*. In other words, just because *Mdl*'s behavior is correct when dealing with data from $D_E$ does not imply that the behavior will be correct when dealing with data from $D_{tot}$.

Generalizing Formulae (13), (14), and (15), we can describe the behavior of *Mdl* when dealing with $D_{tot}$ as follows:

$$D_{tot} = D_S \cup D_F \wedge D_S \cap D_F \neq \emptyset \tag{17}$$

$$D_S, S_D \vdash R \tag{18}$$

$$D_F, S_D \nvdash R \tag{19}$$

However, in general, we are not able to compute

$$acc = \frac{|D_S|}{|D_S|+|D_F|} = \frac{|D_S|}{|D_{tot}|} \tag{20}$$

because |$D_{tot}$| is usually too large a number to make such computation feasible.

*B. Managing the evolution of AIIS*

Suppose we derived from $D_E$ a model *Mdl* that has accuracy $acc_0$ according to Formula (16). We then deployed *Mdl* (together with *M* and *P*) and started using it in production. In the first month of usage, *Mdl* is used on data $D_{m1}$, the second month it is used with $D_{m2}$, etc. Considering the results yielded by *Mdl* every month, it is generally possible to split $D_{mi}$ into $D_{Si}$ and $D_{Fi}$, such that

$$D_{Si}, M, P, Mdl \vdash R$$

$$D_{Fi}, M, P, Mdl \nvdash R$$

so that the accuracy at month *i* can be computed as

$$acc_i = \frac{|D_{Si}|}{|D_{Si}| + |D_{Fi}|}$$

Note that in some cases $acc_i$ can be computed as soon as the month is over: for instance, the number of games lost by the chess playing AIIS is immediately known. In other cases, $acc_i$ becomes known after some delay: for instance, the number of incorrect diagnoses, especially false negatives, is possibly known several months after the diagnosis has been produced.

However, we know that the environment evolves, so that the characteristics of data $D_{mi}$ evolve as well. This evolution can make *Mdl* progressively less capable of achieving *R*. This is quite natural, and it is not a peculiarity of AIIS: tools developed for a given environment can become less effective, or even useless, when the environment changes. By monitoring $acc_i$ we can notice when the accuracy starts decreasing, and when it goes below the acceptability threshold. When this occurs, we can derive a new *Mdl*, by also using the most recent $D_{mi}$ data.

We could also decide to periodically renew *Mdl*, without waiting for accuracy to decrease. In fact, the availability of new data at the end of every considered period makes it possible to build a new *Mdl* that, being trained on a larger set of data, is expected to provide better performance.

*C. Security*

The *M*, *P*, *Mdl* ensemble is subject to the usual security requirements, and could be the object of "traditional" security threats. However, in addition to the well-known security problems that affect most software applications, AIIS suffer from specific problems.

The behavior of an AIIS depends largely on the *Mdl* it embeds. Therefore, an attacker could change the behavior of an AIIS to his/her advantage by affecting the behavior of the *Mdl*. There are at least two strategies that an attacker could adopt to bend the behavior of a *Mdl*.

A first strategy consists in exploiting the features of an existing *Mdl*. Knowing the *Mdl*, and knowing that the results yielded by the AIIS are usually trusted, because in general human users are not able to tell if those results are correct, an attacker could supply the system with input data that he/she knows will induce the *Mdl* to yield an incorrect result. An example of this type of attack concerns AIIS used for detecting attacks over computer networks. These AIIS are trained to recognize communication traffic patterns that are used by attackers. On the contrary, they consider other patterns safe. In these conditions, an attacker may possibly conceive a network attack that shows a safe traffic pattern, so that the attack goes undetected by the AIIS. In other words, the attacker has to find a pattern that belong to $D_F$, and is a false negative. This may turn out to be quite hard, because false negatives are usually expensive, therefore it is likely that only *Mdls* that produce very few false negatives are promoted to production.

A second strategy consists in polluting the data used to derive *Mdl*, so that the resulting *Mdl* behaves as desired by the attacker. This strategy can be seen as a complement of the previous one: the first strategy works only if there is a weak point in the *Mdl*. Via this second strategy, the attacker acts to make sure that *Mdl* will have an exploitable weak point.

Since the *Mdl* is derived from data collected from the environment, an attacker could induce the collection of data that are meant to change the behavior of *Mdl* as desired. As an example, suppose that the attacker wants to "instruct" a chess playing AIIS to perform a very weak move in a given chessboard configuration, so that whoever manages to reach that configuration can take advantage of the AIIS's weak move and, most likely, win the game. To this end, the attacker may inject in the training data a conspicuous set of game recordings where in the given chessboard configuration a player makes the weak move, and eventually wins the game. In this way, the learning algorithm could be induced to believe that the weak move is actually a winning move.

Let $D_O$ be the set of "clean" data that would be used to derive *Mdl* in normal conditions. Let $D_H$ be the hackered data that the attacker manages to introduce in $D_E$, so that

$$D_E = D_O \cup D_H$$

Based on $D_E$, *Mdl* is derived so that

$$D_E, M, P, Mdl, \vdash R$$

with an acceptably high accuracy, as discussed in the previous section. The important point here is that during data analysis and model building the effects of $D_H$ go unnoticed, because they are mixed up with the effects of $D_O$. As long as the accuracy of the *M*, *P*, *Mdl* ensemble is acceptable, it is likely that nobody notices the existence and effects of $D_H$.

In production, we have that $D_F$ is affected by $D_H$: e.g., the AIIS will lose a game that proceeded as described in $D_H$ until a certain chessboard configuration is reached, because its *Mdl* outputs a weak move in the considered situation.

## V. RELATED WORK

Differences between traditional software and AIIS have been noted and described by several researchers and practitioners.

Polyzotis et al. [7] observed that the accuracy of a ML model is deeply tied to the data that it is trained on.

Based on the experience in developing data-centric infrastructure for a production ML platform at Google, the authors highlight some relevant challenges. Specifically, they focus on data management challenges, concerning data understanding, data validation and cleaning, and data preparation. The paper by Polyzotis et al. addresses the processes that are responsible to produce *Mdls*, while recognizing the importance of $D_E$.

Amershi et al. [8] reported about the development of AIIS at Microsoft. They observed a nine-stage workflow process, in which preexisting, well-evolved engineering processes were merged. They also observed best practices to address AIIS-specific challenges. They identified some aspects of the AI domain that make it fundamentally different from traditional software application domains: among these "AI-specific" aspects, one concerns discovering and managing the data needed for ML applications, therefore the importance of $D_E$ and the dependence of *Mdl* on $D_E$ was stressed again.

Ozkaya [9] discussed what is really different in engineering AIIS with respect to traditional software. A first significant difference highlighted by Ozkaya concerned the difficulty to specify AIIS upfront: "*What changes in engineering AI-enabled systems, however, is the fact that uncertainty is their dominant characteristic. In particular, learning from data and the discovery process introduced with ML-modeling activities both introduce many uncertainties.*" As we showed, this problem can be re-stated in terms of the proposed notation: we are able to provide a description of *E* (the behavior of the environment) and *R* (what we would like to achieve), but we do not know how to specify *S* until we 1) get $D_E$, 2) build a suitable *Mdl* based on $D_E$, 3) split $D_E$ into $D_{ES}$ and $P_{EF}$, and 4) evaluate the performance of *Mdl*, hence the satisfaction of functional requirements.

Another issue mentioned by Ozkaya concerns the verification of AIIS correctness: "*Verification challenges are inevitably exacerbated in AI-enabled systems given their inherent uncertainty.*" Also in this case, our proposal can help: we make clear that requirements are not satisfied in all cases, and that we could and should redefine the concept of functional correctness based on the quantification of requirements satisfaction.

Bosch et al. [10] proposed a research agenda for AIIS engineering. Based on the analysis of several case studies, they learned that new, structured engineering approaches are required to build and evolve AIIS. They sketch a research agenda for AI engineering that provides an overview of the key engineering challenges surrounding ML solutions and an overview of open items thatneed to be addressed by the research community at large. Bosch et al. identified a few AI engineering strategic foci. Among these, data quality management stresses the importance of $D_E$ to derive high-quality *Mdls*; Model performance stresses both the importance of performance (as discussed in Section IV-A) and general quality at tributes (like efficiency) of the *M*, *P*, *Mdl* ensemble.

## VI. CONCLUSIONS

Differences between traditional software and AIIS have been noted and described by several researchers and practitioners. However, the lack of a shared set of concepts and a common notation made discussions and descriptions rather imprecise and somewhat ambiguous. Consider for instance the difficulty to specify AIIS upfront, as described by Ozkaya [9] (see Section V). Ozkaya concludes that "*AI specifications are specifications of problems, not systems.*" The notation by Jackson [1] that we propose to use lets us address these issues clearly and precisely.

With the proposed notation, the "problem" is described by $R$, both for traditional software and for AIIS. Sometimes $R$ is well-defined: for instance, the desired AIIS has to be able to perform a given diagnostic task with at least the same accuracy as a given reference set of doctors. However, as described in Section III, the fact that we succeed in developing a system that satisfies $R$ depends on the available data $D_E$ and so does the extent to which $R$ is satisfied (see Section IV): this the uncertainty mentioned by Ozkaya.

Unfortunately, there are also cases in which even requirements $R$ are known only in a very broad and abstract way. In industry, the ever increasing availability of data induces the demand for "opportunistic" AIIS. For instance, let us consider the following request, which is a quite representative example: "we have a lot of data concerning the operation of a given machine (mechanical, electric and digital parameters, operation speed, quality measures of the product, etc.): can we exploit these data to get an AIIS that improves operations?" Note that in this question, the "operation improvement" is unspecified: it may consist of decreasing the set-up time of machines, increasing the quality of products, decreasing the amount of energy needed to operate the machine, decrease the need for maintenance, etc. In these cases, there is a double uncertainty, since 1) data $D_E$ determine the *Mdl* we can derive as part of the implementation of a system that behaves as $S_D$, thus satisfying $R$, and 2) data determine $R$ itself, since $R$ depends on $S_D$, which in its turn depends on $D_E$.

So, uncertainty does not only concern the extent to which we will be able to satisfy the requirement, but also the very content and object of requirements themselves. This is clearly a huge difference with respect to traditional software development, where we never get requests such as "please, take these data and build some program that uses them to do whatever could possibly be useful." In other words, traditional software development proceeds from requirements to programs. Given $R$, we con conceive a machine that behaves according to specifications $S$, which satisfy $R$. Then, we can build a hardware-software machine that implements $S$ (note that is the logical flow of operations, which can be arranged in different types of process models). With AIIS, the situation is very different: given a broad description of $R$, we can conceive $S$, but we are not at all sure that such $S$ (more specifically, its *Mdl* component) can be built using the available data $D_E$. In the opposite direction, given $D_E$, we can derive $S_D$ and a *Mdl* that supports $S_D$, and finally discover to what extent $R$ we can satisfied.

In conclusion, in this paper, we proposed to use the notation introduced by Jackson [1] to describe AIIS and their peculiarities. We showed that in this way it seems possible to write descriptions that help improve the comprehension of AIIS and the rigor of the discussions concerning AIIS development, operations, and maintenance.

In our future work, we plan to use the proposed notation and concepts to describe processes that are suitable for addressing the specific characteristics of AIIS, while preserving our ability to deal with the non AI parts of the systems, which could be relatively large and complex, as in most traditional software systems. Also, we will investigate the introduction of a comprehensive quality model specifically tailored for AIIS, by identifying the specific relevant qualities (including functional correctness and security that we addressed in this paper) and their relationships.

## ACKNOWLEDGMENT

This work has been partially supported by the "Fondo di ricerca d'Ateneo" of the Università degli Studi dell'Insubria.